# Nonlinear effects for three-terminal heat engine and refrigerator

Rongqian Wang[1], Jincheng Lu[1], Chen Wang[2,+], and Jian-Hua Jiang[1,*]

*[1]College of Physics, Optoelectronics and Energy, & Collaborative Innovation Center of Suzhou Nano Science and Technology,*

*Soochow University, 1 Shizi Street, Suzhou 215006, China*

*[2]Department of Physics, Hangzhou Dianzi University, Hangzhou, Zhejiang 310018, China*

+ wangchen@hdu.edu.cn

* jianhuajiang@suda.edu.cn

The three-terminal heat device that consists of an electronic cavity and couples to a heat bath is studied both as a heat engine and as a refrigerator. We investigate the characteristic performance in the linear and nonlinear regime for both setups. It is our focus here to analyze how the efficiency of the heat engine and coefficient of performance of the refrigerator are affected by the nonlinear transport. With such considerations, the maximum efficiency and power are then optimized for various energy levels, temperatures and other parameters.

The thermoelectricity is a direct conversion of heat into electrical energy, or electricity into heat[1-3]. Thermoelectric devices at nanoscales with high efficiency and power have attracted much attention due to the enhanced figure of merit and energy conversion efficiency[3-18], compared with bulk materials. However, to date, thermoelectric materials still have a very low efficiency in converting heat into electrical work and deliver only moderate power. Recently, many strategies have been proposed to further improve the efficiency and coefficient of performance in thermoelectric nanodevices. In particular, two-terminal geometries, which uses quantum dots as typically efficient energy filters, have been considered and shown great improvements of the thermoelectric performance[19-26]. Another important strategy to make the energy conversion efficient is to add a thermal terminal to the conventional two-terminal geometries. In the past years, there has been a growing interest in three-terminal thermoelectric setups[27-33]. Intriguingly, the third terminal makes it possible to control the heat and electric current separately, which helps to reduce parasitic heat leakage and leads to a higher efficiency.

A quantum heat engine, as a representative three-terminal setup, generates power from the heat flow between hot and cold reservoirs. In the existing literature, a variety of three-terminal heat engines have been proposed and investigated in the linear response regime[7,33-37], which is valid when the thermodynamic biases (e.g., temperature difference and voltage bias) are small[38]. Specifically, a thermal engine working in the linear response regime was investigated and a general formalism for the efficiency at maximum power was unraveled[39]. Meanwhile, the upper bound of the efficiency at maximum output power for all thermodynamic system was proved to be 50%. However, in realistic the nanoscale devices often operate in the nonlinear regime due to their small sizes, instead of the linear response limit. Later, theoretical and experimental studies on nonlinear thermoelectric transports have been performed in two-terminal or elastic thermoelectric devices[40-42]. However, the nonlinear transport effects in these studies turn out to be marginally. Recently, Jiang and Imry have showed that nonlinear effects can dramatically enhance the efficiency and power of three-terminal quantum heat engines, while for two-terminal devices such improvements are significantly reduced. The underlying physics is revealed as due to drastic increase of the density of phonons (or other bosons), which assists the inelastic transport in three-terminal quantum heat engines[43]. On the other hand, a quantum refrigerator as a reversed operation compared to the heat engine, pumps heat from a cold to a hot bath by consuming power. However, the coefficient of performance (COP) of three-terminal refrigerators, has never been studied in the nonlinear transport regime.

In this work, we study the nonlinear effects on the COP of three-terminal refrigerators and heat engines based on a set-up that the cavity serves as a reactor to transfer electrons, which is efficiently thermalized by the thermal bath. Specifically, as an electron enters the cavity with an energy $E_l$, it absorbs the energy gap $\Delta E = E_r - E_l$ from the cavity, and leaves the cavity afterwards, which finally gives rise to a net electrical current. This process is different from the set-up in Ref. 43, in which electrons jump from the left quantum dot (with low energy level) to the right quantum dot (with high energy level) both via coherent electron tunneling and phonon directly mediated scattering, without including a cavity. Moreover, the three-terminal set-up is quite similar to the device in the work of Edwards *et al.*[17], where the central cavity is sandwiched between two electrodes and coupled to an electronical load by a tunneling junction. However, the cavity in Ref. 17 is to be cooled, whereas it performs as an auxiliary component in our set-up.

In the following, we calculate the energy-conversion efficiency, electric power, electric current and the transport heat in both linear and nonlinear regime. For the heat engine, the nonlinear efficiency and output power are enhanced compared to the linear transport. While for the refrigerator, the nonlinear efficiency and cooling power are reduced to nearly half of the linear ones. We optimize the maximum efficiency and power by tuning the energy levels, temperatures, and other parameters. Our results show that nonlinear effects can improve the maximum efficiency of the heat engine to 25% of the Carnot efficiency (with parasitic heat leakage included) and the maximum power to more than an order of the linear counterpart.

## Model and Formalism

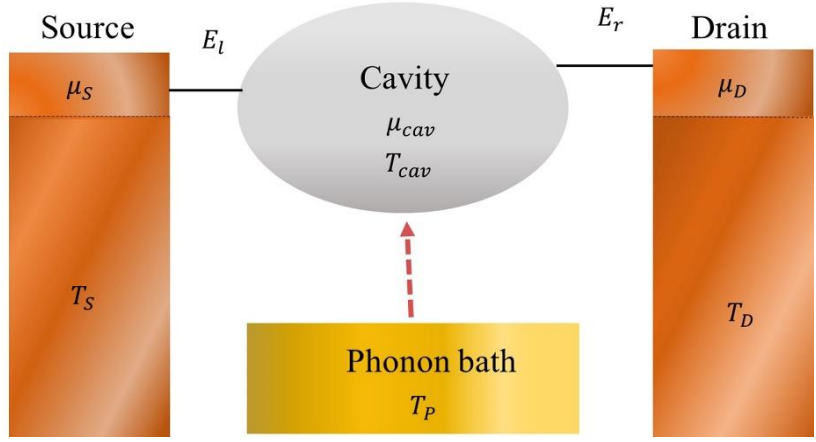

**Figure. 1.** Schematic view of a three-terminal thermoelectric system. The three-terminal device is composed by two electronic reservoirs (characterized by their temperatures $T_{ele}$ and chemical potential $\mu_{S(D)}$) and a phonon bath, which is held at temperature $T_{Ph}$. The central cavity, which is thermalized by the phonon bath, is connected to two electrodes via two quantum dots at energy $E_{l(r)}$.

The three-terminal thermoelectric device we consider is illustrated in Fig. 1. The left (right) quantum dot is directly in contact with the left (right) electronic reservoir. It describes an electron leaves the source into the QD1, and hops to the QD2 through the cavity, which is thermalized by the phonon bath. Then, it finally tunnels into the drain. The electronic reservoirs, $i = S$, $D$, are characterized by the Fermi-Dirac function $f_i(E) = 1/\left[\exp\left(\frac{E - \mu_i}{k_B T_i}\right) + 1\right]$. We assume that strong electron-electron and electron-phonon interactions relax the electron energies as they enter and leave the electronic cavity. Hence, the occupation function of the cavity can also be described by the Fermi-Dirac function, $f_{cav}(E) = \frac{1}{\left[\exp\left(\frac{E - \mu_{cav}}{k_B T_{cav}}\right) + 1\right]}$, completely characterized by a chemical potential $\mu_{cav}$ and temperature $T_{cav}$. [27] To reach steady state, the cavity must exchange energy with the phonon bath (denoted by a brown arrow in Fig. 1). We assume that the thermal conduction between the phonon bath and the cavity is efficient so that the temperature gradient is considerably small. In this way, one can approximately treat that $T_{cav} = T_{Ph}$.

The Hamiltonian of the electronic system is described as

$$H = H_S + H_D + H_{cav} + H_{QD} + H_{int}, \tag{1}$$

where $H_S$, $H_D$, $H_{cav}$ and $H_{QD}$ are the Hamiltonians of the source, the drain, the cavity and the QD, respectively. Specifically, $H_\alpha = \sum_{\vec{k}} \varepsilon_{\vec{k}} c_{\vec{k},\alpha}^\dagger c_{\vec{k},\alpha}$, where $\alpha = S, D, cav$ denotes the source, drain, and the electron cavity. $c_{\vec{k},\alpha}^\dagger (c_{\vec{k},\alpha})$ creates(annihilates) one electron in the $\alpha th$ bath, and the electron energy is $\varepsilon_{\vec{k}} = \frac{\hbar^2 k^2}{2m^*}$, with $m^*$ the effective mass and $k$ the wave vector of the charge carrier. The Hamiltonian of quantum dots is shown as

$$H_{QD} = \sum_i E_i d_i^\dagger d_i, \tag{2}$$

where $d_i^\dagger d_i$ denotes particle number operators for the dots, respectively, with $i = L, R$ representing left and right quantum dots. The interaction Hamiltonian which describes the hybridization of the QD states with the states in the source, drain and cavity is given by

$$H_{int} = \sum_{\alpha=S,D,cav}\sum_{\vec{k},i} V_{i,\alpha,\vec{k}}\, c_{\vec{k},\alpha}^\dagger d_i + \text{H. c.,} \tag{3}$$

with $V_{i,\alpha,\vec{k}}$ the inteaction strength between the ith dot and $\alpha$th bath. To capture the transport effects, we apply a bias $V = \mu/e$ to this system. The chemical potential of the source and the drain are set anti-symmetrically, i.e. , $\mu_S = -\mu_D = \mu/2$.

Generally, the electron current through the left (right) electrode into cavity can be evaluated by Landauer-Buttiker formalism[44]

$$I_{e,j} = \frac{2e}{h}\int dE \mathcal{T}_j(E)[f_j(E) - f_{cav}(E)] \tag{4}$$

where $j = l, r$, $h$ is the Plank's constant and $\mathcal{T}_j(E)$ is the energy dependent transmission function. Here for simplicity, we neglect the contributions of the dot-dot coupling and lead-lead interaction. Then, by using the nonequilibrium Green's function approach[1], we obtain the non-interacting transmission function as a Lorentzian shape[45]

$$\mathcal{T}_j(E) = \frac{\Gamma_{j1}(E)\Gamma_{j2}(E)}{(E-E_j)^2 + \left[\frac{\Gamma_{j1}(E)+\Gamma_{j2}(E)}{4}\right]^2}, \tag{5}$$

where energy-dependent coupling strengths of the quantum dot to the source (drain) and the cavity are

$$\Gamma_{j1}(E) = \frac{2\pi}{\hbar}\sum_{\vec{k}}\left|V_{S(D),\vec{k}}\right|^2\delta(E - \varepsilon_{\vec{k}}), \tag{6a}$$

$$\Gamma_{j2}(E) = \frac{2\pi}{\hbar}\sum_{\vec{k}}\left|V_{cav,\vec{k}}\right|^2\delta(E - \varepsilon_{\vec{k}}), \tag{6b}$$

and $f_j(E)(j = l, r)$ denotes the particle occupation of the left and right electrodes. To conserve the electron current, the chemical potential of the cavity can be determined as

$$I_{e,l} + I_{e,r} = 0. \tag{7}$$

While for the heat current flowing from the source (drain) to the cavity, $I_{Q,j}$ has two contributions, i.e.,

$$I_{Q,j} = I_{Qe,j} + I_{Qp,j}, \tag{8}$$

where

$$I_{Qe,j} = \frac{2}{h}\int dE(E - \mu_j)\mathcal{T}_j(E)[f_j(E) - f_{cav}(E)], \tag{9a}$$

$$I_{Qp,j} = \int_0^\infty \frac{dE}{h} E \mathcal{T}_{ph}(E)\left[n_B\left(\frac{E}{k_B T_j}\right) - n_B\left(\frac{E}{T_{cav}}\right)\right], \tag{9b}$$

are the electrical and phonon heat current flowing from the source (drain) to the cavity, respectively. Here $n_B(x) = \frac{1}{(e^{\beta x}-1)}, \left(\beta = \frac{1}{k_B T}\right)$ is the Bose-Einstein distribution function. $\mathcal{T}_j(E)$ is the non-interacting transmission probability for electrons, and $\mathcal{T}_{ph}(E)$ is the ideal transmission function for phonons. To consider the low-frequency phnonons, which dominate the steady state behavior, it can be expressed as $\mathcal{T}_{ph}(E) = \alpha\Theta(E_{cut} - E)$ with $\alpha$ a dimensionless constant and $E_{cut}$ the cut-off energy of the phonons. Phonons with energy lower than $E_{cut}$ can spring out of the bath and interact with electrons, while the higher energy phonons are bounded in the bath. Moreover, the conservation of energy results in

$$\dot{E}_S + \dot{E}_D + \dot{E}_P = 0, \tag{10}$$

where $\dot{E}_i = \dot{Q}_i + \mu_i \dot{N}_i \ (i = S, \ D, \ P)$ stands for the total energy in the source (drain), and $\dot{E}_P = \dot{Q}_P$ is the energy flow in the phonon bath. $\dot{N}_{S(D)} = -I_{e,l(r)}/e \ (e < 0)$ represents the particle current flowing into the source (drain). Combining Eq. (4), (5) and (7), we obtain the heat injected into the system from the phonon bath as

$$I_{Q,P} = -\dot{Q}_P = -I_{Q,l} - \mu_S \frac{I_{e,l}}{e} - I_{Q,r} - \mu_D \frac{I_{e,r}}{e} \tag{11}$$

Then, the total entropy production rate[46] of the system is contributed from three currents and corresponding thermodynamic forces

$$\dot{S}_{tot} = \frac{\dot{Q}_S}{T_S} + \frac{\dot{Q}_D}{T_D} + \frac{\dot{Q}_P}{T_P} = \sum_i I_i A_i. \tag{12}$$

Specifically, the first term is the electron current driven by chemical potential bias between electrodes, shown as

$$I_e = -e\dot{N}_S, \ A_e = \frac{\mu_S - \mu_D}{e} \left( \frac{1}{2T_S} + \frac{1}{2T_D} \right). \tag{13}$$

The second term is the energy current of electrons under the temperature bias of two electrodes

$$I_{Q,e} = \frac{1}{2} (\dot{Q}_D - \dot{Q}_S), \ A_{Q,e} = \frac{1}{T_D} - \frac{1}{T_S}.$$

While the third term originates from the heat flow of phonons from the thermal bath

$$I_{Q,P} \ = -\dot{Q}_P, \ A_{Q,P} = \frac{1}{2T_S} + \frac{1}{2T_D} - \frac{1}{T_P},$$

In our work, we set $T_S = T_D = T_{ele}$, $T_P = T_{Ph}$. Thus, we can simplify the forces as

$$A_e = \frac{V}{T_{ele}}, \ V = \frac{\mu_S - \mu_D}{e}, \tag{14}$$

$$A_{Q,e} = 0, \ A_{Q,P} = \frac{1}{T_{ele}} - \frac{1}{T_{Ph}}.$$

In the linear regime ($\Delta T \ll T_{ele}, T_{Ph}, \ \Delta V \ll \frac{\mu_S}{e}, \frac{\mu_D}{e}$), the thermodynamic fluxes and forces follow the Onsager relations[47]

$$I_i = \sum_j M_{ij} A_j, \tag{15}$$

where only the lowest order of thermodynamics biases need to be considered. Specifically,

$$I_e = M_{11} A_e + M_{12} A_{Q,P}, \tag{16a}$$

$$I_{Q,P} = M_{21} A_e + M_{22} A_{Q,P}, \tag{16b}$$

where the coefficients $M_{ij}$ are the Onsager coefficients satisfying the reciprocal relation $M_{12} = M_{21}$. The second law of thermodynamics requires that[48]

$$M_{11} \geq 0, \ M_{22} \geq 0, \ M_{11} M_{22} - M_{12}^2 \geq 0. \tag{17}$$

Generally, the linear transport coefficients can be obtained by calculating the ratios between currents and affinities in the linear response regime with very small voltage bias and temperature difference[43].

**Nonlinear transport effects on a three-terminal heat engine**

The three-terminal device can be tuned into a heat engine by setting $T_{ele} = T_c$, $T_{Ph} = T_h$. The heat engine has the ability to convert the absorbed heat into the electric power, which is expressed as $P_{out} = -I_e V$. Here $I_e$ is the net electrical current through the system as the charge conservation implies $I_e = I_{e,l} = -I_{e,r}$. The energy-conversion efficiency is then defined by the ratio of the injected heat and the output power

$$\eta = \frac{P_{out}}{Q_{in}}, \tag{18}$$

where $Q_{in} = I_{Q,P}$ is the heat current flowing from the phonon bath due to the temperature difference between the electrode and the heat bath. Considering the physical significance, the efficiency is well-defined only in the regime with $P > 0$ and $Q_{in} > 0$. Consequently, the Carnot efficiency for heat engine is defined by the temperature of the electrode and the phonon bath

$$\eta_C = 1 - \frac{T_c}{T_h} \tag{19}$$

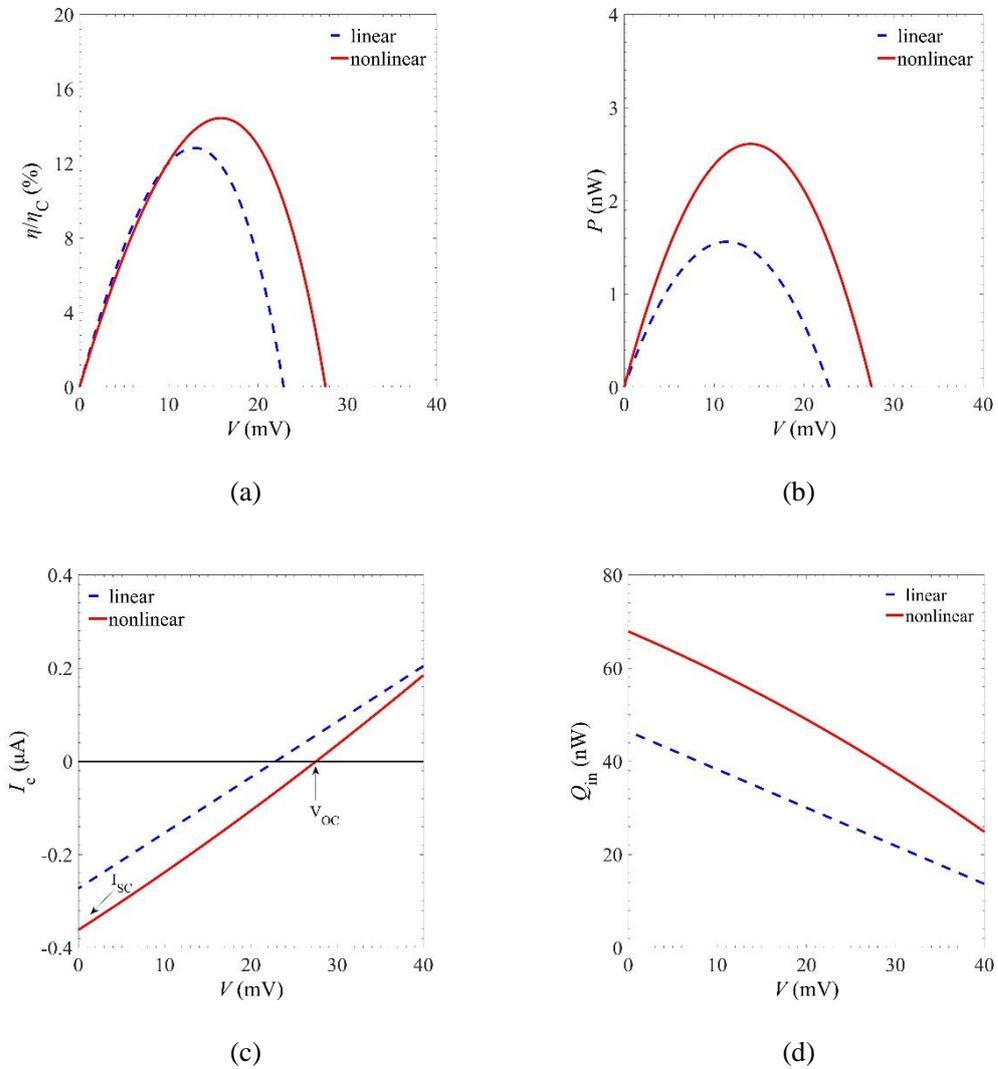

(a)

(b)

(c)

(d)

**Figure. 2.** (Color online) Performance of the three-terminal heat engine. **(a)** Energy efficiency $\eta$ in units of the Carnot efficiency $\eta_C$ and **(b)** output power $P$ as functions of voltage $V$ (in units of mV). **(c)** The electrical and **(d)** heat currents as functions of voltage $V$. Parameters are $\Gamma = 30$ meV, $E_{cut} = 100$ meV, $\alpha = 0.1$, $k_B T_c = 30$ meV, $k_B T_h = 45$ meV and $E_l = -E_r = -60$ meV.

We firstly analyze the efficiency and output power for a three-terminal heat engine in both linear and nonlinear regimes. At a fixed temperature $T_h = 1.5T_c$, the nonlinear transport yields significant improvement of the maximum efficiency and power, as shown in Fig. 2(a) and 2(b). The maximum efficiency under small voltage bias is 12.8% of the Carnot efficiency, while the full calculation (including the nonlinear transport effect) reaches 14.7% of the Carnot efficiency, which is 1.2 times of the linear counterpart. Moreover, the maximum power in linear regime is 1.6 nW, whereas it increases to 2.6 nW in the nonlinear regime. Hence, we conclude that the nonlinear effect significantly enhances the thermoelectric performance.

To better understand the enhancing mechanism of the maximum efficiency and power, we then investigate how the electrical and heat currents are affected by the nonlinear transport. Fig. 2(c) shows that the electrical current is considerably enhanced due to the nonlinear effect. The short-circuit current $I_{sc}$ (the electrical current at zero voltage $V = 0$) is increased by 1.3 times. We can interpret this from current formulas Eq. (14) and (16a), the contribution of linear $I_{sc}$ only comes from $A_{Q,P}$, the temperature difference of electrode and heat bath. While the nonlinear effect not only contains contribution of temperature difference, but also originates from the multilevel channels of transported electrons. The open-circuit voltage $V_{oc}$ is the voltage at which $I = 0$, indicating a dynamic equilibrium between the source, drain and the cavity, which generates the zero output power and efficiency in Fig. 2(a) and (b). The nonlinear $V_{oc}$ raises to 1.3 times of that in the linear response regime. The product of the short-circuit current and open-circuit bias gives a nonlinear output power $P$ more than 1.6 times as large as the one in the linear regime, which agrees well with the improvement of the maximum power. Besides, we also examine how the input heat $Q_{in}$ is affected by the nonlinear transport effect. Fig. 2(d) reveals that the maximum heat input at $V = 0$ increases to about 1.5 times as large as that obtained in the linear limit. Hence, the increase of the output power exceeds that of the input heat, which clearly unravels the improvement of the maximum efficiency.

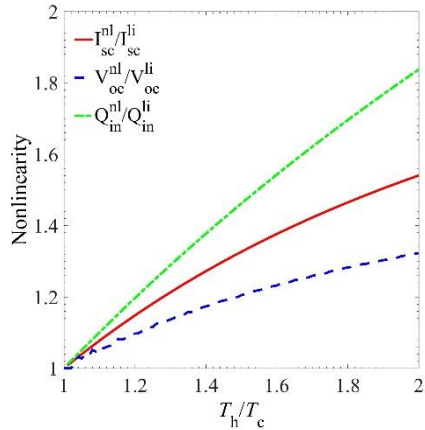

**Figure. 3.** (Color online) The ratios of the short-circuit current, open-circuit voltage and the heat current in the nonlinear and linear regimes, as functions of $T_\mathrm{h}/T_\mathrm{c}$. Parameters are $\Gamma = 30$ meV, $k_\mathrm{B}T_\mathrm{c} = 30$ meV and $E_l = -E_r = -60$ meV.

Then, we turn to explore the nonlinear effects in the thermoelectric transport. We plot the short-circuit electrical, heat currents and the open-circuit voltage as functions of $T_\mathrm{h}/T_\mathrm{c}$, by fixing $T_\mathrm{c}$. As presented in Fig. 3, it is exhibited that $I_\mathrm{sc}^\mathrm{nl}/I_\mathrm{sc}^\mathrm{li}$、 $V_\mathrm{oc}^\mathrm{nl}/V_\mathrm{oc}^\mathrm{li}$ and $Q_\mathrm{in}^\mathrm{nl}/Q_\mathrm{in}^\mathrm{li}$ all increase rapidly when $T_\mathrm{h}$ is raised. Specially for $T_\mathrm{h} \geq 1.2T_\mathrm{c}$, the "nonlinear" currents and voltage are more than 10% larger than the linear ones. Such enhancement is mainly due to the multichannel induced electron transport.

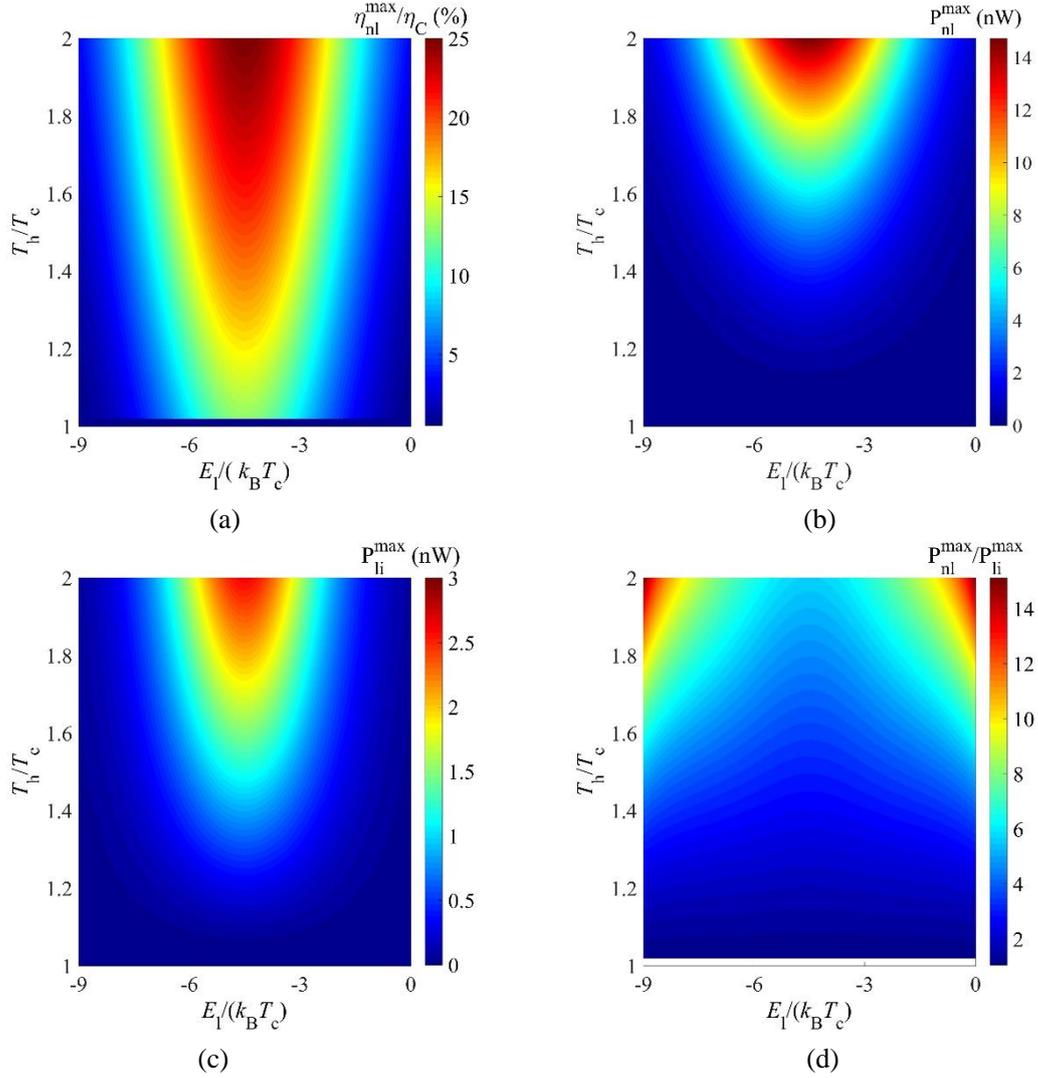

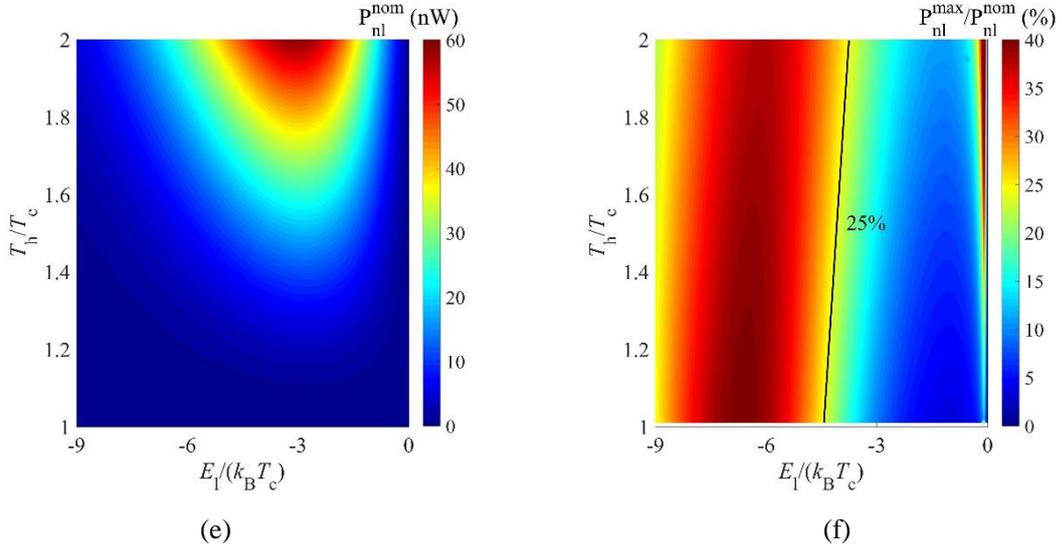

(e)                                    (f)

**Figure. 4.** (Color online) **(a)** Energy efficiency $\eta_{nl}^{max}/\eta_C$ **(b)** output power $P_{nl}^{max}$ **(c)** $P_{li}^{max}$ and **(d)** their ratio $P_{nl}^{max}/P_{li}^{max}$ **(e)** the nominal power $P_{nl}^{nom} \equiv I_{sc}V_{oc}$ and **(f)** the filling factor (the ratio of the maximum power and the nominal power), as functions of the QD energy $E_l$ (in units of $k_B T_c$), for $E_r = E_l + 9k_B T_c$. Parameters are $\Gamma = 30$ meV, $E_{cut} = 100$ meV, $\alpha = 0.1$ and $k_B T_c = 30$ meV.

Furthermore, we study the effect of nonlinear transport on thermoelectric energy by modulating temperatures and QD energies. The ratio of the maximum efficiency over the Carnot efficiency $\eta_{nl}^{max}/\eta_C$ in Fig.4(a) and the maximum power in Fig. 4(b) and 4(c) are exhibited, with $T_P = T_h$ and $E_r - E_l = 9k_B T_c$. The optimal efficiency in linear regime is well defined by $\frac{\eta_{li}^{max}}{\eta_C} = \frac{\sqrt{ZT+1}-1}{\sqrt{ZT+1}+1}$, where $ZT = \frac{M_{12}^2}{M_{11}M_{22}-M_{12}^2}$ is the proverbial figure of merit, which shows its independence on temperature ratio $T_h/T_c$. The optimal value of $\frac{\eta_{li}^{max}}{\eta_C}$ is about 12% at $E_l = -E_r = -4.5k_B T_c$. While the ratio $\eta_{nl}^{max}/\eta_C$ can directly reflect enhancement of the nonlinear transport effect. Fig. 4(a) shows that the $\eta_{nl}^{max}/\eta_C$ can increase from 5% to 25%, which is more than twice of the linear counterpart. Fig. 4(d) demonstrates that the enhancement factor $P_{nl}^{max}/P_{li}^{max}$ can be as large as 14, which shows acute dependence on the temperature ratio $T_h/T_c$. Remarkable improvement can be reached, with $\frac{\eta_{nl}^{max}}{\eta_C} \sim 10\%, P_{nl}^{max}/P_{li}^{max} \sim 20\%$, even when the temperature ratio is rather small ($\sim 1.02$).

Finally, we calculate the filling factor of the three-terminal heat engine to evaluate how much power can the device extracts, compared to the theoretical maximum power, namely the nominal power, which is defined as $P_{nl}^{nom} \equiv I_{sc}V_{oc}$. From Fig. 4(f), we can find that the nominal power $P_{nl}^{nom}$ strongly depends on the temperature ratio and the QD energy. The maximum value appears at $T_h = 2T_c$ and $E_l \cong -3.5k_B T_c$. On the contrary, the filling factor in the nonlinear regime (defined as $P_{nl}^{max}/P_{nl}^{nom}$) presented in Fig. 4(e) varies slightly with the temperature ratio. It optimizes in the energy range of $-9k_B T_c < E_l < -4.5k_B T_c$. The filling factor in the linear regime $P_{li}^{max}/P_{li}^{nom}$ is a constant of 1/4 over all temperature range and different energy levels. Therefore, the nonlinear effect enhances the useful power by more than one order-of-magnitude compared to the linear limit.

**Nonlinear transport effects on a three-terminal refrigerator**

The three-terminal device can be tuned to be a refrigerator, by exchanging temperatures of the electrode and the phonon bath, i.e., $T_{S(D)} = T_h$, $T_{ph} = T_c$, with $T_h > T_c$. Then, the phonon bath can be cooled, and heat $Q_{out}$ is transferred to the cavity. Here, we use the invested work as the chief power supplier, $P_{in} = IV$. The cooling efficiency is defined by the ratio of the cooling heat $Q_{out}$ and the input power $P_{in}$

$$\eta = \frac{Q_{out}}{P_{in}} \tag{17}$$

and the Carnot efficiency for which the process is reversible is given by

$$\eta_C = \frac{T_c}{T_h - T_c} \tag{18}$$

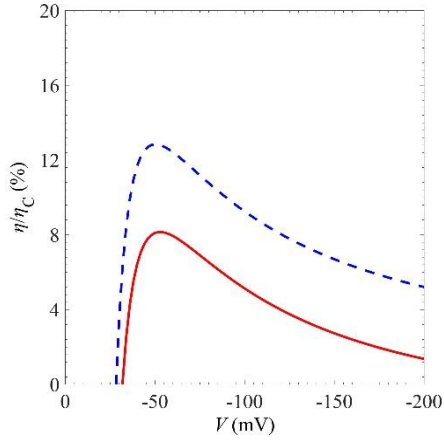

(a)

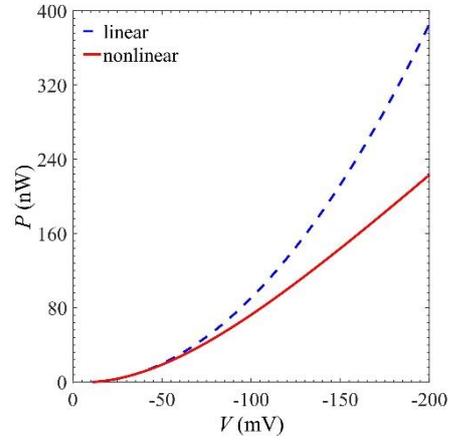

(b)

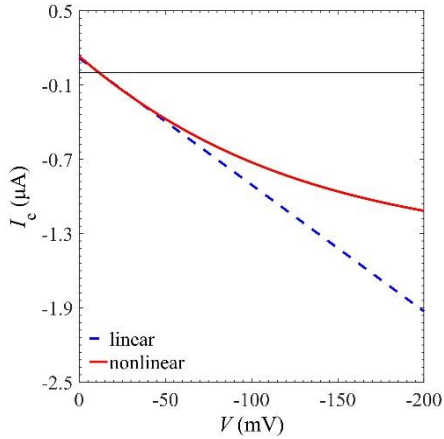

(c)

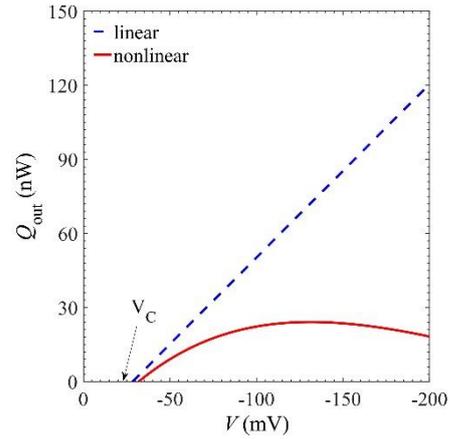

(d)

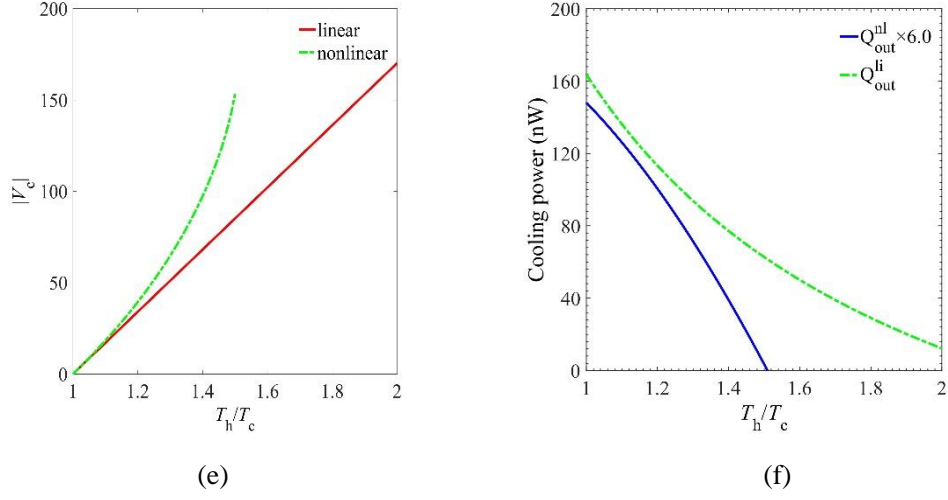

<div align="center">(e)                            (f)</div>

**Figure. 5.** (Color online) Performance of the three-terminal refrigerator. **(a)** COP $\varepsilon$ in the unit of Carnot efficiency $\varepsilon_C$ and **(b)** input power $P$ as a function of voltage $V$ (in units of mV); **(c)** Electrical and **(d)** heat currents with $k_B T_c = 30$ meV, $k_B T_h = 35$ meV; **(e)** Threshold bias and **(f)** the cooling power for various $T_h/T_c$. The other parameters are $\Gamma = 30$ meV, $E_{cut} = 100$ meV, $\alpha = 0.1$ and $E_l = -E_r = -60$ meV.

We first study the COP and input power for a three-terminal refrigerator by the same method as done at Eqs. (11), (14), (16) and (18). The temperatures are selected as $T_h = 405$ K and $T_c = 347$ K. Fig. 5(a) indicates that the nonlinear transport effect reduces the coefficient of performance, with maximum COP $\eta^{nl}/\eta_C$ three-fifths of the linear one, which is contrary to the Fig. 2(a) (the heat engine case). The electric power injected into the system is shown in Fig. 5(b). $P_{max}^{nl}$ is 41.7% lower than $P_{max}^{li}$, which indicates that under the same voltage bias, the nonlinear transport effect consumes much less electric power.

To find out why the cooling efficiency is reduced in the nonlinear regime, we study how the electrical and heat currents are affected by the nonlinear transport, as presented in Fig. 5(c) and (d). As the Eq.16(a) indicates, the negative current increases with the voltage bias in the linear regime. While the nonlinear one, which contains the contribution of the multichannel transport, does not increase as fast as the linear one. Specifically, the negative maximum electric current via linear-approximation calculation is 1.71 times as large as the full calculation one, which is in accordance with the input power. Fig. 5(d) shows that the maximum cooling heat with nonlinear effect firstly increases and then decreases with the voltage bias. While the linear one grows continually with the bias. The trend of the nonlinear cooling heat is determined by the Fermi-Dirac distribution factor $f_j(E) - f_{cav}(E)$ $j = S, D$, which saturates at large voltage bias. Moreover, lower production lagging behind the consumption leads to the deterioration of the cooling efficiency. The value of the bias at which the cooling heat current starts to flow is called the threshold $V_c$. Fig. 5(d) indicates that the "working regime" (the voltage range over which cooling is possible) of the linear effect is slightly extended than the nonlinear effect. This is in consistent with Fig. 5(a) that the cooling efficiencies appear only when the bias exceed a certain value.

Fig. 5(e) makes it clear that how the threshold bias is determined by the temperature ratio $T_h/T_c$. The $V_C$ via linear approximation increases with the temperature ratio, while the nonlinear one end abruptly when $T_h$ reaches 1.5 times of $T_c$. To find out the reason, we examine the behavior of the cooling power under the same circumstances. As Fig. 5(f) presents that the $Q_{\text{out}}^{\text{li}}$ remains positive even when $T_{\text{h}} \approx 2T_{\text{c}}$, as Eq. 16(b) shows. However, the $Q_{\text{out}}^{\text{nl}}$ reduces to zero when $T_{\text{h}} \approx 1.5T_{\text{c}}$, which indicates an energy balance between source and drain. We can conclude that the cooling power for nonlinear transport effect is limited to low voltage bias.

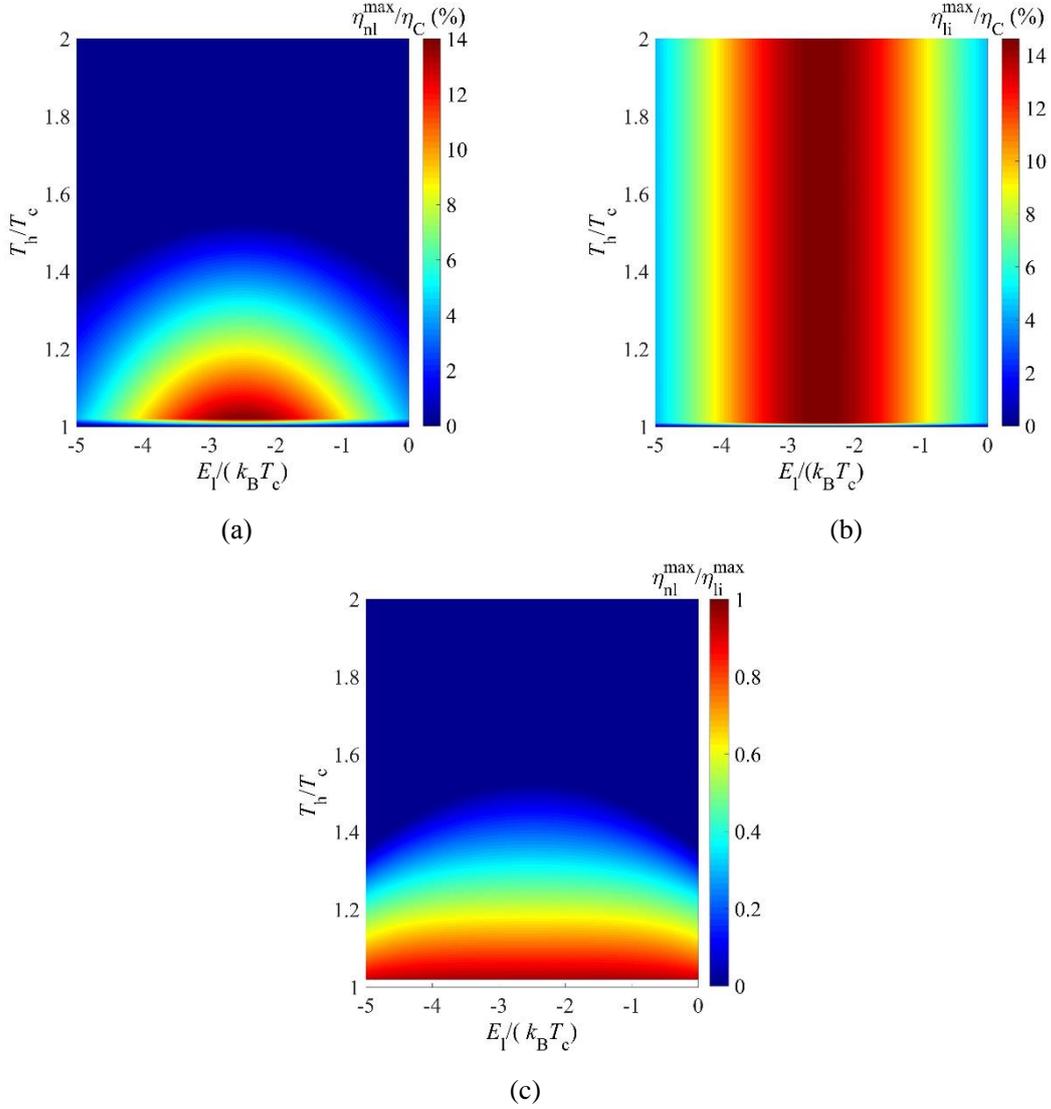

**Figure. 6.** (Color online) **(a)** Energy efficiency $\eta_{\text{nl}}^{\text{max}}/\eta_{\text{C}}$, **(b)** $\eta_{\text{li}}^{\text{max}}/\eta_{\text{C}}$ and **(c)** their ratio $\eta_{\text{nl}}^{\text{max}}/\eta_{\text{li}}^{\text{max}}$ vs QD energy $E_{\text{l}}$ (in the unit of $k_{\text{B}}T_{\text{c}}$) for $E_{\text{r}} = E_{\text{l}} + 5k_{\text{B}}T_{\text{c}}$. Parameters are $\Gamma = 30$ meV, $E_{\text{cut}} = 100$ meV, $\alpha = 0.1$ and $k_{\text{B}}T_{\text{c}} = 30$ meV.

We then turn to analyze the comprehensive effect of the dot energy and the temperature, shown in Fig.6. We set the QD energy difference $E_{\text{r}} - E_{\text{l}} = 5k_{\text{B}}T_{\text{c}}$ and vary the temperature from $T_{\text{c}}$ to $2T_{\text{c}}$. It is found that the optimization value of $\eta_{\text{nl}}^{\text{max}}/\eta_{C}$ appears around $E_{\text{l}} = -E_{\text{r}} \cong -2.5k_{\text{B}}T_{\text{c}}$, which testifies that the "particle-hole symmetric" configuration is also best for a cooling machine. Fig. 6(a) also presents that the cooling efficiency diminishes gradually with the increasing temperature ratio $T_{\text{h}}/T_{\text{c}}$, where Fig. 5(f) may account for this diverting phenomenon. The cooling heat decreases with increasing $T_{\text{h}}$ at fixed energy, and reaches zero at $T_{\text{h}} \approx 1.5T_{\text{c}}$, which leads to the zero cooling efficiency. Fig. 6(b) gives the cooling efficiency enhancement factor $\eta_{\text{nl}}^{\max}/\eta_{\text{li}}^{\max}$ under the same parameters as in Fig. 6(a). It is exhibited that the maximum nonlinear efficiency $\eta_{\text{nl}}^{\max}$ can reach 90% of the linear one when the temperature bias is very small.

## Conclusions

In summary, we have investigated the influence of nonlinear response of three-terminal setup on the thermoelectric performance, including efficiency, power, electric and heat currents. We find that the nonlinear effect can significantly improve the performance of the three-terminal heat engine. When the temperatures of the electrodes and phonon bath are interchanged, the device turns to be a refrigerator. Unlike the heat engine, the nonlinear transport effect considerably reduces the efficiency and cooling power of the three-terminal refrigerator. We also optimize the efficiency and power at different parameters, in which the optimal values can be reached as the device becomes "particle-hole symmetric", with the dot energy $E_{\text{l}} = -E_{\text{r}}$. From the practical view, three-terminal thermoelectric devices have already been fabricated in experiments where the electron cavity is made of GaAs/Al$_x$Ga$_{1-x}$As heterostructure, and NiCr/Au gates were patterned on the GaAs/Al$_x$Ga$_{1-x}$As heterostructure surface by using electron beam lithography.[17,49-51] The heat bath can be implemented by insulating substrates. With these advancements, our study can serve to improve the understanding on three-terminal thermoelectric energy conversion and provide insights for the design and operation principles of nanoscale thermoelectric devices.

## Acknowledgments

R.W., J.L., and J.H.J are supported by the National Natural Science Foundation of China (No.11675116) and the Soochow University. C.W. is supported by the National Natural Science Foundation of China (No.11704093).

## Author contributions statement

R.W. and J.L. performed theoretical calculation and theory. J.H.J and C.W. guided the research. All authors contributed to the analysis of the results and the composition of the manuscript.

## Additional information

**Competing financial interests:** The authors declare no competing financial interests.

## Corresponding authors

Correspondence should be addressed to J.H.J (jianhuajiang@suda.edu.cn) or C.W (wangchen@hdu.edu.cn )